\documentclass{appolb}
\usepackage{epsfig}

\begin{document}
\title{Structural Hamiltonian of the international trade network\thanks{Presented at Summer Solstice 2011 Conference on Discrete Models of Complex
Systems, Turku, Finland.}%
}
\author{Agata Fronczak
\address{Faculty of Physics, Warsaw University of Technology,
Koszykowa 75, \\PL-00-662 Warsaw, Poland}}
\maketitle
\begin{abstract}
It is common wisdom that no nation is an isolated economic island. All nations participate in the global economy and are linked together through trade and finance. Here we analyze international trade network (ITN), being the network of import-export relationships between countries. We show that in each year over the analyzed period of 50 years (since 1950) the network is a typical representative of the ensemble of maximally random networks. Structural Hamiltonians characterizing binary and weighted versions of ITN are formulated and discussed. In particular, given binary representation of ITN (\ie binary network of trade channels) we show that the network of partnership in trade is well described by the configuration model. We also show that in the weighted version of ITN,  bilateral trade volumes (\ie directed connections which represent trade/money flows between countries) are only characterized by the product of the trading countries' GDPs, like in the famous gravity model of trade. \end{abstract}

\section{Introduction}

In this contribution, we continue the analysis of structural properties of the international trade network that has been initiated in Ref.~\cite{2011_Fronczak}. Although the new results given here are mainly related to binary network analysis of ITN, the weighted network approach (as originally presented in \cite{2011_Fronczak}) is also described for a clear exposition of the whole theory and for better reception of ideas gathered in the last section entitled \emph{Discussion and perspectives}.

Thus, in this paper, we study international trade \cite{2003_PRE_Serrano, 2003_PhysA_Li, 2004a_PRL_Garlaschelli, 2005_PhysA_Garlaschelli, 2008_JStatMech_Bhattacharya, 2009_PRE_Fagiolo, 2010_PRE_Barigozzi, 2010_JEvolEcon_Fagiolo} from a complex network perspective \cite{2002_RMP_Albert, 2003_SIAM_Newman, 2003_book_Dorogovtsev, 2008_book_Barrat,2009_Science_nets}. The knowledge of topological properties of this network and its evolution over time is not only important per se (\eg, because it enhances our descriptive knowledge of the stylized facts pertaining to the ITN), but it may also be relevant to a better explanation of macroeconomic dynamics \cite{2007_JIntBus_Kali, 2009_Science_Schweitzer, 2009_AdvCS_Schweitzer}. In particular, it has been suggested that the analysis of ITN may help in recognizing the pattern of economic interdependencies responsible for the propagation of crises across countries \cite{2007_JEconInter_Serrano, 2007_EPJB_Reichardt, 2008_AdvCS_Reyes, 2010_EconInq_Kali, 2010_NewJPhys_Garas} and that it can be used to explain the role of international trade in spurring the efficiency of economic recovery after recession \cite{2007_EPJB_Garlaschelli, 2009_ChinPL_Wen}. The approach to international trade described in this article is a significant step towards clarifying the above-mentioned macroeconomic issues.

We use quantitative and numerical (data-driven) methods originating from statistical mechanics to describe the behavior of ITN. We analyze a set of year-by-year trade relationships between all countries of the world, covering the time interval $1950-2000$ \cite{2002_JRes_Gleditsch}. Although the total number of countries and the overall economic conditions influencing the network change over the course of the period, in each year ITN is shown to be a typical representative of the ensemble in which every network, $G$, is assigned the probability \cite{2004_PRE_Park, 2009_PRL_Garlaschelli} $P(G)\propto e^{-H(G)}$, where $H(G)$ plays the role of network structural Hamiltonian. The statement holds true for both: binary and weighted versions of ITN. In the former case, the network Hamiltonian is given by $H(G)=\sum_i\theta^{(k)}_ik_i$, where $k_i$ is the number of trade partners of a country $i$, $\theta^{(k)}_i\propto \ln x_i$ is the external parameter describing geopolitical conditions influencing international trade of this country, with $x_i$ corresponding to the country's GDP. In the later (\ie weighted) case, we argue that the global trade is described by $H(G)=\sum_{ij}\theta^{(w)}_{ij}w_{ij}$, where $w_{ij}$ represents the volume of trade between two countries, $i$ and $j$, with $\theta^{(w)}_{ij}\propto(x_ix_j)^{-1}$ being the field parameter conjugated to this trade connection and $x_ix_j$ corresponding to the product of the GDPs of the trade partners.

In the following we show that the two ensembles which describe structural properties of ITN are characterized by factorizable partition functions that can be calculated exactly. We also argue that behind the descriptive power of our approach (which is confirmed in a number of tests consisting in comparison of GDP-driven Monte Carlo simulations of the trade network with real data on ITN), it also reveals interesting predictive abilities, providing (as reported in \cite{2011_Fronczak}), through fluctuation-response theorems, valuable insights into general rules governing time evolution of the global trade.

\section{Statistical mechanics of networks}

In the last decade, research on complex networks has become pervasive in many disciplines, ranging from mathematics to the social, economic, and biological sciences \cite{2009_Science_nets}. Network structures, different from regular lattices, have also attracted much interest in the community of physicists. Various concepts of statistical physics have been used to understand how real-world networks change over time, how this affects their structural properties, and what the interplay is between their topology and dynamics. A number of models have been proposed to embody the structural characteristics and to explain the functional properties of the considered networks. The models can roughly be divided into two classes: static (equilibrium) and evolving (causal, non-equilibrium). The second class of causal networks encompasses, in particular, the famous Barab\'{a}si-Albert model \cite{1999_Science_Barabasi}, whereas exponential random graphs  \cite{2004_PRE_Park}, exploited in this study to analyze ITN, belong to the first class of static networks.

Exponential random graphs are ensemble models. The methodology behind the models directly follows that behind the maximum entropy school of thermodynamics \cite{1957_PR_Jaynes}. In order to correctly define an ensemble of networks, one has to specify a set of networks, $\mathcal{G}=\{G\}$, that one wants to study. Next, one has to decide what constraints should be imposed on the ensemble. For example, the constraints may be encouraged by the properties of real networks. (In our case they will be guided by the properties of ITN.) Then one specifies probability distribution, $P(G)$, over the ensemble, which consists in maximization of the Shannon entropy, $S=-\sum_{G\in\mathcal{G}}P(G)\ln P(G)$, subject to the given constraints. The procedure leads to the Boltzmann-like probability distribution
 \begin{equation}\label{PG}
P(G)=\frac{e^{-H(G)}}{Z},
\end{equation}
where
\begin{equation}\label{ZG}
Z=\sum_{G\in\mathcal{G}}e^{-H(G)}
\end{equation}
stands for the partition function, whereas $H(G)$ is called the network Hamiltonian.

In general, the network Hamiltonian can be written in the following form
\begin{equation}\label{HG}
H(G)=\sum_{i} \theta_i A_i(G),
\end{equation}
where $\{A_i(G)\}$ represents the set of free parameters of the ensemble upon which the relevant constraints act, and $\{\theta_i\}$ is the set of fields conjugated to these parameters (like energy and the inverse temperature in the canonical ensemble). The set $\{\theta_i\}$ completely determines average values of the free parameters, $\{\langle A_i\rangle\}$, which can be calculated as appropriate derivatives of the partition function
\begin{equation}\label{meanA}
\langle A_i\rangle=\sum_{G\in\mathcal{G}}A_i(G)P(G)=-\frac{\partial\ln Z(\{\theta_i\})}{\partial\theta_i}.
\end{equation}

The well-known example of exponential labelled random graphs is the ensemble of networks with a given sequence of expected degrees \cite{2004_PRE_Park}, $\{\langle k_i\rangle\}$. In this ensemble, the set $\mathcal{G}$ consists of all simple graphs with a fixed number of nodes, $N$. A simple graph has, at most, one undirected link between any pair of vertices, and it does not contain self-loops connecting vertices to themselves. Such a graph is completely described by the symmetric adjacency matrix, $\mathbf{A}$, whose elements $a_{ij}$ equal either $1$ or $0$, depending on whether there is a connection between $i$ and $j$ or not. The network Hamiltonian characterizing this ensemble can be written as
\begin{equation}\label{Hk}
H(G)=\sum_{i=1}^N \theta^{(k)}_i k_i(G)=\sum_{i<j}(\theta^{(k)}_i+\theta^{(k)}_j)a_{ij},
\end{equation}
where $k_i=\sum_ja_{ij}$ and the coefficients $\{\theta^{(k)}_i\}$ represent a kind of hidden variable (or fitness parameter) assigned to nodes \cite{2002_PRL_Caldarelli, 2003_PRE_Boguna}. (In what follows, since it does not result in confusion, we will write $\theta_i$ instead of $\theta^{(k)}_i$.) The partition function of the model is given by
\begin{equation}\label{Zk}
Z(\{\theta_i\})=\prod_{i<j}(1+e^{-(\theta_i+\theta_j)}).
\end{equation}
For the analysis performed in subsequent sections of this article, it is significant that Eq.~(\ref{Zk}) allows one to rewrite the probability of a network in the considered ensemble, \ie Eq.~(\ref{PG}), in the following form
\begin{equation}\label{PGk}
P(G)=\prod_{i<j}p_{ij}^{a_{ij}}(1-p_{ij})^{1-a_{ij}},
\end{equation}
where
\begin{equation}\label{pijk}
p_{ij}=\frac{e^{-(\theta_i+\theta_j)}}{1+e^{-(\theta_i+\theta_j)}}.
\end{equation}
This, in turn, enables one to identify $p_{ij}$ with the probability that nodes $i$ and $j$ are connected.

\section{Hamiltonian of the international trade network}

\subsection{Data used and notation}

The results reported in this work are based on the empirical analysis of expanded trade data collected by K. S. Gleditsch \cite{2002_JRes_Gleditsch}. The data set \cite{dataset} contains, for each world country in the period $1950-2000$, the detailed list of bilateral import and export volumes. The data on the population size of each country and its GDP \emph{per capita} has been taken from the Penn World Tables (PWT) \cite{PWT}.

The trade data are employed to build a sequence of matrices, $\mathbf{W}(t)$, corresponding to snapshots of weighted directed ITN in the consecutive years, $t=1950,\dots,2000$. In the network, each country is represented by a node and the direction of links follows that of wealth flow. The entry, $w_{ij}(t)$, of the trade matrix, $\mathbf{W}(t)$, represents the weight of the directed connection. From the point of view of the country denoted by $i$, $w_{ij}(t)$ refers to the volume of export to $j$, while, from the point of view of the country labeled by $j$, it is seen as the volume of import from $i$. Precisely due to differences in reporting procedures between countries, when analyzing the data one often encounters small deviations between exports from $i$ to $j$ and imports from $i$ to $j$. To overcome the problem, in our analysis we define $w_{ij}(t)$ as the arithmetic average of the two values.

In this article, apart from trade matrices, which contain complete but often excessively detailed information about ITN, we also use several other quantities that make theoretical description of the network possible. In particular, to characterize the economic performance of a country we use its total GDP value, $x_i(t)$. To get the whole set of total GDPs, $\{x_i(t)\}$, we simply multiply the GDP \emph{per capita} by the population size of each country. Furthermore, to describe the intensity of the trade relationships of a country we define the so-called out-strength, $s_i^{out}(t)$, and in-strength, $s_i^{in}(t)$, of the corresponding node \cite{2004_PNAS_Barrat}. The quantities are calculated as the total weight of connections (outgoing and incoming, respectively) that are attached to the node, \ie
\begin{equation}\label{si}
s_i^{out}(t)=\sum_{j}w_{ij}(t)\;\;\;\mbox{and}\;\;\;s_i^{in}(t)=\sum_{j}w_{ji}(t),
\end{equation}
and they represent total volumes of export and import of the considered country in a given year, $t$.

\begin{figure}
\centerline{\epsfig{file=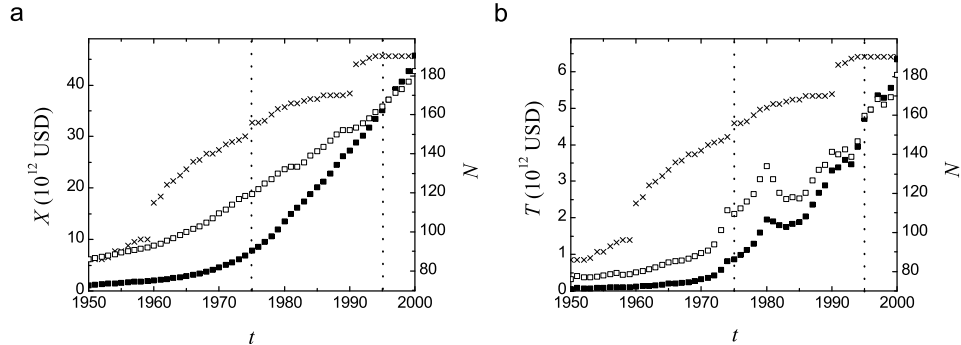,width=\columnwidth,angle=0}}
\caption{\label{fig1} \textbf{Global economic indicators characterizing temporal evolution of ITN.} \textbf{a,} Global GDP, Eq.~(\ref{Xt}), vs. time. \textbf{b,} World's trade volume, Eq.~(\ref{Xt}), vs. time. Filled and open squares in panels (\textbf{a}) and (\textbf{b}) correspond to values of $X$ and $T$ expressed in current and $1996$ U.S. dollars, respectively. In both panels, crosses represent the number of world countries vs. time. The vertical dashed lines in the figure mark two years, 1975 and 1995, whose trade characteristics are used in Figures \ref{fig2} and \ref{fig3} to illustrate topological properties of ITN.}
\end{figure}

Finally, we would like to stress that both GDP data, $\{x_i(t)\}$, and trade matrices, $\mathbf{W}(t)$, used in this study are given in millions of contemporary U.S. dollars (that is, in terms of the value of one U.S. dollar in the reported year, $t$). To factor out the effects of inflation and to allow for a direct comparison between snapshots of ITN in different years, one usually expresses the data in standard reference money units (\eg in 1996 U.S. dollars \cite{2007_EPJB_Garlaschelli}). In our approach, however, the disturbing effects mentioned are ruled out in a natural way by the fact that whenever the variables, $x_i(t)$ or $w_{ij}(t)$, are used in the calculations, they are intrinsically divided by the normalization constant that equals the sum of all variables of a given type, \ie $\sum_ix_i(t)$ or $\sum_{i,j}w_{ij}(t)$, respectively. (In what follows, whenever there is no confusion we will often omit explicit time dependence of the quantities). As a byproduct, the above observation allows one to present the main results of this article in a very concise way, with the help of relative quantities defined as follows:
\begin{equation}\label{xiit}
\xi_i=\frac{x_i}{X},\;\;\;\;\;\sigma_i^{(\dots)}=\frac{s_i^{(\dots)}}{T}\;\;\;\mbox{and}\;\;\; v_{ij}=\frac{w_{ij}}{T},
\end{equation}
where $X$ stands for the sum of the total GDPs of all the countries, whereas $T$ represents the world's trade volume (see Figure~\ref{fig1}), \ie
\begin{equation}\label{Xt}
X=\sum_ix_i\;\;\;\mbox{and}\;\;\;T=\sum_{i,j}w_{ij}.
\end{equation}

\subsection{Binary network approach}

The first contributions dealing with international trade from a complex network perspective used a binary-network approach, in which one has assumed that a (possibly directed) link between any two countries is either present or not, depending on whether the trade volume that it carries is larger than a given threshold \cite{2003_PRE_Serrano, 2003_PhysA_Li, 2004a_PRL_Garlaschelli}.

With reference to this line of research we would like to highlight the paper by Garlaschelli and Loffredo (2004), published in \emph{Physical Review Letters} \cite{2004a_PRL_Garlaschelli}. In the paper, the authors used the same real-world data to analyze an unweighted and undirected version of ITN, \ie a network of partnership in trade. They have shown that the total GDP of a country, $x_i$, can be identified with the fitness variable \cite{2002_PRL_Caldarelli} that, once a form of the probability of trade connection between two countries is introduced, completely determines the expected structural properties of this network. The assumed expression for the probability was given as follows
\begin{equation}\label{pijGL}
p_{ij}=\frac{\delta x_ix_j}{1+\delta x_ix_j},
\end{equation}
where the parameter $\delta>0$ only depends the total number of the world's trade channels. The value of $\delta$ can be estimated from the relation below (see also Figure 3b in \cite{2005_PhysA_Garlaschelli}):
\begin{equation}\label{new1}
L\simeq\sum_{i>j}p_{ij},
\end{equation}
where $L=(\sum_ik_i)/2$ represents the total number of trade channels in a given snapshot of ITN, while $k_i$ is the number of trade partners of a country. Remarkably, Eq.~(\ref{pijGL}) allows the creation of a statistical ensemble of networks whose average structural properties are in complete agreement with the ones observed in binary representation of ITN (\cf Figures 2, 3 in \cite{2004a_PRL_Garlaschelli}, and 4, 5 in \cite{2007_EPJB_Garlaschelli}).

Given the explicit expression for the probability that two countries trade goods with each other, it is interesting to note that this probability, \ie Eq.~(\ref{pijGL}), has the identical form as the corresponding probability in the ensemble of networks with a given degree sequence, cf. Eq.~(\ref{pijk}). This in turn implies that ITN viewed as a simple graph is characterized by the well-known network Hamiltonian,~Eq.~(\ref{Hk}), which, when written in the variables specific for trade, takes the form
\begin{equation}\label{new2}
H(G)=-\sum_{i=1}^N \ln(x_i\sqrt{\delta}) k_i=-\sum_{i=1}^N\ln(x_i) k_i- \ln(\delta) L.
\end{equation}

Now, from the perspective of exponential random graphs, ITN appears to be a network in which every country tries to keep an optimal (from its own viewpoint) number of trade connections. However, using Eq.~(\ref{meanA}), one can show that the expected connectivity of a node, $\langle k_i\rangle=\sum_{j}p_{ij}$, does not only depend on $x_i$, \ie on the country itself. It also depends on global economic performance, which is expressed by the parameter $\delta$, and also on the GDPs of other countries.

\subsection{Weighted network analysis}

Since ITN viewed as a binary network is a typical representative of exponential random graphs characterized by a well-defined Hamiltonian, one can expect that the same holds true for the weighted version of the network. To verify this conjecture, we start by considering the most general ensemble of directed weighted networks, which is described by the following Hamiltonian
\begin{equation}\label{Hw}
H(G)=\sum_i\sum_{j\neq i}\theta^{(w)}_{ij}w_{ij},
\end{equation}
with a separate parameter $\theta^{(w)}_{ij}$ coupling to each weighted connection. (In what follows, we will write $\theta_{ij}$ instead of $\theta^{(w)}_{ij}$.) Our aim is to check whether the Hamiltonian is correct and, if so, how the parameters $\{\theta_{ij}\}$ depend on different indicators characterizing the global economy. In order to do this, we first examine the ensemble as it stands.

Thus, given that $w_{ij}$ is a real number greater than $0$ (as is true for trade volumes), the partition function, Eq.~(\ref{ZG}), of this ensemble can be written as
\begin{equation} \label{Zw}
Z(\{\theta_{ij}\})=\prod_i\prod_{j\neq i}\int_0^\infty e^{-\theta_{ij}w_{ij}} dw_{ij}=\prod_i\prod_{j\neq i}\frac{1}{\theta_{ij}}.
\end{equation}
This allows us to rewrite the probability of a network, Eq.~(\ref{PG}), in a way similar to what has been done in the case of undirected binary networks with an expected degree sequence, \ie
\begin{equation} \label{PGw}
P(G)=\prod_i\prod_{j\neq i}p_{ij}(w_{ij}),
\end{equation}
where
\begin{equation} \label{pijw}
p_{ij}(w_{ij})=e^{-\theta_{ij}w_{ij}}\theta_{ij}
\end{equation}
is the probability that there is a directed link of weight $w_{ij}$ from $i$ to $j$. The expression for $p_{ij}(w_{ij})$  that we arrive at is the exponential distribution. Its mean value, that is given by
\begin{equation} \label{meanw}
\langle w_{ij}\rangle=\frac{1}{\theta_{ij}},
\end{equation}
can be used to calculate the average values of a node's out- and in-strength \begin{equation}\label{meansi}
\langle s_i^{out}\rangle=\sum_{j\neq i}\langle w_{ij}\rangle=\sum_{j\neq i}\frac{1}{\theta_{ij}} \;\;\;\;\;\mbox{and}\;\;\;\;\;\langle s_i^{in}\rangle=\sum_{j\neq i}\frac{1}{\theta_{ji}}.
\end{equation}

\begin{figure}
\centerline{\epsfig{file=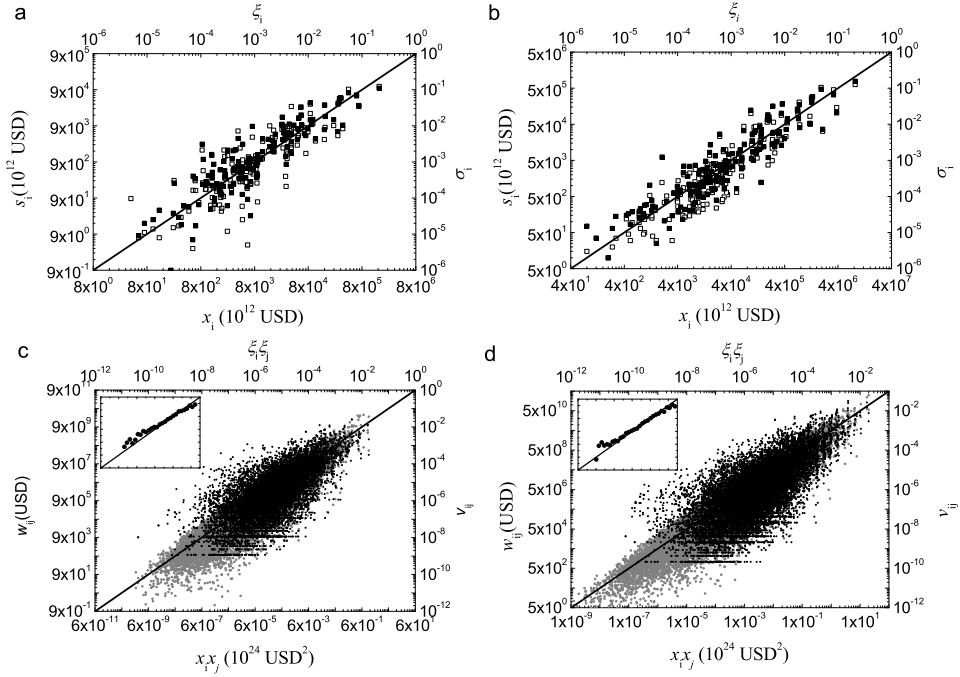,width=\columnwidth,angle=0}}
\caption{\label{fig2} \textbf{Weighted network approach. Structural properties of two different snapshots of ITN.} (The left column refers to the year 1975, while the right column to the year 1995. The data are shown in two ways, by using regular and relative quantities, cf.~Eq.~(\ref{xiit}).) \textbf{a, b,} Total import and export volumes of all world countries in a given year vs. GDP (filled and open points, respectively) and their comparison with the expected values described by Eqs.~(\ref{siITN}) and (\ref{sigmaITN}) (solid lines).  \textbf{c, d,} Bilateral trade flows versus the product of the trading countries' GDPs (points) as compared with their theoretical prediction based on Eqs.~(\ref{meanw}) and (\ref{omegaITN}) (lines). Black points correspond to real data, while gray points represent trade volumes obtained from GDP-driven Monte Carlo simulations. Since trade flows smaller than a given threshold are rarely specified in economic reports (in particular, the considered data set \cite{dataset} does not contain trade volumes smaller than 1000 USD), the clouds of black points cover smaller areas than the ones corresponding to numerical simulations. The effect of unreported exports/imports is also perceptible in the insets of both panels, which show the relationship between the average trade volume and the product of GDPs obtained with logarithmic binning of the latter. In the insets, points represent real data, while the solid lines stand for their theoretical prediction.}
\end{figure}

At this stage one can start to make comparisons of theoretical predictions with the empirical data on international trade. With good reason, it is convenient to begin by putting together Eq.~(\ref{meansi}) and the corresponding empirical relations (see Fig.~\ref{fig2},~a and b):
\begin{equation}\label{siITN}
\langle s_i^{out}\rangle=Ax_i\;\;\;\;\;\mbox{and}\;\;\;\;\;\langle s_i^{in}\rangle=Ax_i,
\end{equation}
where $A$ is the time-dependent parameter having the same value for both out- and in-strength of the nodes. Analyzing the expressions, one finds that the simplest way to merge the theoretical approach with real data is to assume a multiplicative form of the parameter $\theta_{ij}$, \ie
\begin{equation}\label{thetaij}
\theta_{ij}=\theta_i\theta_j,
\end{equation}
where $\theta_i$ and $\theta_j$ represent some single-node characteristics controlling for the potential ability of the two nodes to be connected. One should note that the symmetric expression for $\theta_{ij}$, Eq.~(\ref{thetaij}), is consistent with observations made by other authors, showing the symmetric character of bilateral trade relations \cite{2004b_PRL_Garlaschelli}.

To calculate explicit values of all the parameters $\{\theta_i\}$, one just has to insert Eq.~(\ref{thetaij}) into the theoretical formula for $\langle s_i^{out}\rangle$, Eq.~(\ref{meansi}), and then equate the obtained relation to the empirical one, Eq.~(\ref{siITN}). (The analogous calculations can be done for $\langle s_i^{in}\rangle$.) As a result, one gets the expression
\begin{equation}\label{pom1}
\langle s_i^{out}\rangle=\frac{1}{\theta_i}\sum_j\frac{1}{\theta_j}=Ax_i,
\end{equation}
which, when summed over $i$, yields an important relation between theoretical and empirical quantities describing ITN, \ie
\begin{equation}\label{pom2}
T=\left(\sum_i\frac{1}{\theta_i}\right)^2=AX,
\end{equation}
from which it follows that
\begin{equation}\label{thetai}
\theta_{i}=\frac{1}{\sqrt{T}}\frac{X}{x_i}=\frac{1}{\sqrt{T}}\frac{1}{\xi_i},
\end{equation}
and
\begin{equation}\label{thetaij2}
\theta_{ij}=\frac{1}{T}\frac{1}{\xi_i\xi_j},
\end{equation}
where the relative quantity $\xi_i$ has been introduced earlier, in Eq.~(\ref{xiit}).

The two expressions, Eqs.~(\ref{thetai}) and~(\ref{thetaij2}), together with other relative parameters that were defined in Eq.~(\ref{xiit}), can be used to rewrite the most important results of this section in a very concise way. In particular, as described in terms of trade, the average out- and in-strength of a node, Eq.~(\ref{meansi}), when divided by the world's trade volume $T$ turns out to be equal to the country's share in the world's GDP, \ie
\begin{equation}\label{sigmaITN}
\langle\sigma_i^{out}\rangle=\langle\sigma_i^{in}\rangle=\xi_i.
\end{equation}
In a similar fashion, the average weight of a directed connection when divided by $T$, is given by
\begin{equation}\label{omegaITN}
\langle v_{ij}\rangle=\xi_i\xi_j.
\end{equation}
Finally, the structural network Hamiltonian, Eq.~(\ref{Hw}), when written in relative variables has the following form
\begin{equation}\label{Hwr}
H(G)=\sum_{i}\sum_{j\neq i}\theta^*_{ij}v_{ij},
\end{equation}
where
\begin{equation}\label{thetastar}
\theta^*_{ij}=\frac{1}{\xi_i\xi_j}=\frac{1}{\langle v_{ij}\rangle}.
\end{equation}

\subsection{GDP-driven Monte Carlo simulations.}

\begin{figure}
\centerline{\epsfig{file=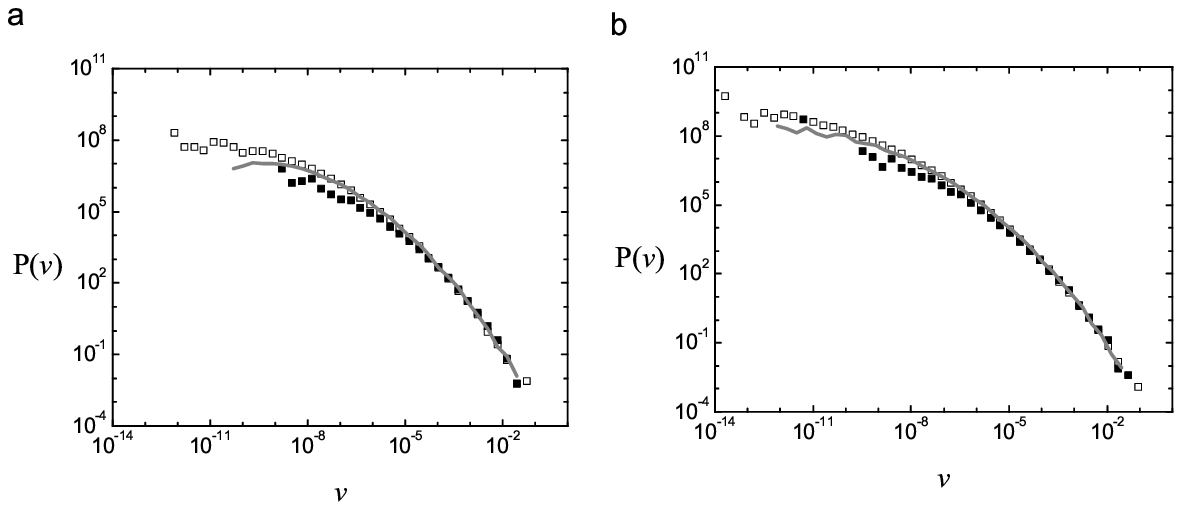,width=\columnwidth,angle=0}}
\caption{\label{fig3} \textbf{Weighted network approach. Distributions of trade volumes.} The left panel \textbf{a,} refers to the year 1975, while the right one \textbf{b,} to the year 1995. Filled and open squares correspond to real and simulated data, respectively. The solid lines represent distributions of expected trade flows which, for each pair of countries in a given year, can be calculated using Eqs.~(\ref{meanw}) or (\ref{omegaITN}).}
\end{figure}

To verify the correctness of the assumed network Hamiltonians, \ie Eqs. (\ref{Hw}) and (\ref{Hwr}), a series of data-driven Monte Carlo simulations employing the Metropolis algorithm has been performed. The obtained results are shown in Figures~\ref{fig2} and \ref{fig3}.

In particular, in Figure~\ref{fig2} c and d, the set of all bilateral trade volumes recorded in a given year versus the product of the trading countries' GDPs is compared with the corresponding set of weights of directed connections in a typical network of the considered ensemble. Although the two sets (clouds) of points are quite dispersed, they overlap significantly, and their shape is well-described by Eq.~(\ref{omegaITN}). Moreover, as shown in Figure~\ref{fig3} a and b, the distributions of trade volumes within these clouds fit very well with each other and agree with the distribution of expected trade flows, $P(\langle v_{ij}\rangle)$, testifying in favor of our simple, yet realistic, approach.

\section{Discussion and perspectives}

\subsection{Quasi-static time evolution of ITN and susceptibility of bilateral trade relations}

Extensive comparisons between real data on international trade and its GDP-driven Monte Carlo simulations show that, although the total number of world's countries, $N(t)$, and their GDPs, $\{x_i(t)\}$, change over the analyzed period of $50$ years (\cf Figure~\ref{fig1}), ITN is continuously well-characterized by the same structural Hamiltonians. This means that the time evolution of this network may be considered as a continuous sequence of equilibrium states (\ie quasi-static process) that is yearly sampled by the national reporting procedures. Furthermore, since differences between snapshots of ITN in the consecutive years are rather small, one can expect that they could be described with the help of linear response theory, of which the simplest (but not yet trivial) result is the fluctuation-response theorem applying to equilibrium ensembles.

In the weighted case of ITN characterized by the Hamiltonian given by Eq.~(\ref{Hwr}) the fluctuation-response theorem can be written in the following form:
\begin{equation}\label{frt}
\frac{d\langle v_{ij}\rangle}{\langle v_{ij}\rangle} =\frac{d\xi_{i}}{\xi_{i}}+\frac{d\xi_j}{\xi_j},
\end{equation}
from which it follows that relative changes in normalized (\ie divided by $T$) bilateral trade volumes can be estimated on the basis of changes in the GDP of trade partners. Yearly changes in import/export volumes between different countries prove that the fluctuation-response theorem for ITN derived in this section is correct \cite{2011_Fronczak}.

Relying on Eq.~(\ref{frt}) one can, for example, expect that a decline of, say, $2$ percent in the relative GDP of a country, given that its trade partners do not change their share of the world's GDP, will translate into a similar decline in all its bilateral trade volumes. The example shows that the theorem can be used to make simple predictions about the world-wide diffusion of trade-based economic perturbations.

\subsection{Gravity model of trade}

Inspired by Newton's gravity equation, the gravity model of trade \cite{2004_book_Feenstra, 1979_AmEconRev_Anderson, 1985_RevEconStat_Bergstrand, 2010_JEcon_Fagiolo} states that the volume of trade between two countries is proportional to the product of the sizes of the two countries (\eg their GDP) and the inverse of the distance between them, \ie
\begin{equation}\label{gravity1}
w_{ij}\propto\frac{x_ix_j}{r_{ij}}.
\end{equation}
Comparing the above expression with Eq.~(\ref{omegaITN}), which arises from our treatment of ITN, one finds a close correspondence between the two. Even though the correspondence exposes possible shortcomings of our approach (stemming from the lack of dependence of trade on distance) we use it to highlight the advantages resulting from using exponential random graphs to study ITN, as compared with the famous econometric model of trade.

To begin with, one should note that the gravity model is an ad hoc method based on empirical reasoning and conventions, rather than a thorough theoretical discussion. Unlike the econometric treatment, in view of our approach (originating from statistical mechanics) the observed properties of ITN arise from an optimization process (equivalent to the maximization of entropy), in which every country tries to optimize its trade channels. The fluctuation-response theorem, Eq.~(\ref{frt}), which follows from our approach, also results from this process, whose details are not known. In conclusion, one should stress that it is impossible to derive the theorem from econometric analysis.

\subsection{Partition function of ITN}

It is interesting to note that the partition function characterizing ITN, Eq.~(\ref{Zw}), can be written as a product of separate partition functions associated with each country, \ie
\begin{equation} \label{Zw1}
Z(\{\xi_{i}\},T)=\prod_i\left(\frac{1}{T}\frac{1}{\xi_i^2}\right)^{N-1}=\prod_iZ_i(\xi_i,T),
\end{equation}
where we have used Eq.~(\ref{thetaij2}). (In the binary version of ITN, the factorizability of the partition function holds true as well.) The factorizability of the partition function characterizing international trade implies that, when considered from the perspective of only trade linkages, the countries are independent, which, in turn, suggests that perturbations in international trade are unlikely to trigger big disruptions at the level of the global economy. Although theoretically reasonable, the intriguing findings are not correct enough. They would be correct if the values of GDP were independent of trade and given \emph{a priori}, as we take them in our approach. However, since, in general, a trade surplus (deficit) translates into GDP growth (decline) there is feedback between trade volume and GDP that causes the observed factorizability to lose its recognized meaning.

\subsection{Perspectives}

Having the mathematically tractable yet realistic model of ITN introduced here, and given the quasi-static time evolution of this network, we believe that, apart from the fluctuation-response theorem, Eq.~(\ref{frt}), other well-known results of non-equilibrium statistical physics \cite{2004_PT_Ruelle} may be applied to estimate recession (or economic growth) impact on international trade.

\section{Acknowledgements}$\\$

This work was supported by the Polish Ministry of Science, Grant No. 496/N-COST/2009/0.


\begin{thebibliography}{6}
\bibitem{2011_Fronczak} A. Fronczak, P. Fronczak, \emph{Statistical mechanics of the international trade network}, arXiv:1104.2606v1.
\bibitem{2003_PRE_Serrano} M. \'{A}ngeles Serrano, M. Bogu\~{n}\'{a}, \emph{Phys. Rev. E} \textbf{68} 015101(R) (2003).
\bibitem{2003_PhysA_Li} X. Li, Y. Y. Jin, G. Chen, \emph{Physica A} \textbf{328} 287 (2003).
\bibitem{2004a_PRL_Garlaschelli} D. Garlaschelli, M. Loffredo, \emph{Phys. Rev. Lett.} \textbf{93} 188701 (2004).
\bibitem{2005_PhysA_Garlaschelli} D. Garlaschelli, M. Loffredo, \emph{Physica A} \textbf{355} 138 (2005).
\bibitem{2008_JStatMech_Bhattacharya} K. Bhattacharya, G. Mukherjee, J. Saram\"{a}ki, K. Kaski, S. S. Manna, \emph{J. Stat. Mech.} P02002 (2008).
\bibitem{2009_PRE_Fagiolo} G. Fagiolo, J. Reyes, S. Schiavo, \emph{Phys. Rev. E} \textbf{79} 036115 (2009).
\bibitem{2010_PRE_Barigozzi} M. Barigozzi, G. Fagiolo, D. Garlaschelli, \emph{Phys. Rev. E} \textbf{81} 046104 (2010).
\bibitem{2010_JEvolEcon_Fagiolo} G. Fagiolo, J. Reyes, S. Schiavo, \emph{J. Evolutionary Econ. } \textbf{20} 479 (2010).
\bibitem{2002_RMP_Albert} R. Albert, A.-L. Barab\'{a}si, \emph{Rev. Mod. Phys.} \textbf{74} 47 (2002).
\bibitem{2003_SIAM_Newman} M. E. J. Newman, \emph{SIAM Rev.} \textbf{45} 167 (2003).
\bibitem{2003_book_Dorogovtsev}S. N. Dorogovtsev, J. F. F. Mendes, \emph{Evolution of Networks: From Biological Nets to the Internet and WWW} (Oxford Univ. Press, 2003).
\bibitem{2008_book_Barrat} A. Barrat, M. Barth\'{e}lemy, A. Vespignani, \emph{Dynamical Processes on Complex Networks} (Cambridge Univ. Press, 2008).
\bibitem{2009_Science_nets} \emph{Complex systems and networks. special issue}, \emph{Science} \textbf{325} 357-504 (2009).
\bibitem{2007_JIntBus_Kali} R. Kali, J. Reyes, \emph{J. Int. Bus. Stud.} \textbf{38} 595 (2007).
\bibitem{2009_Science_Schweitzer} F. Schweitzer, G. Fagiolo, D. Sornette, F. Vega-Redondo, A. Vespignani, D. R. White, \emph{Science} \textbf{325} 422 (2009).
\bibitem{2009_AdvCS_Schweitzer} F. Schweitzer, G. Fagiolo, D. Sornette, F. Vega-Redondo, D. R. White, \emph{Adv. Complex Syst.} \textbf{12} 407 (2009).
\bibitem{2007_JEconInter_Serrano} M. \'{A}ngeles Serrano, M. Bogu\~{n}\'{a}, A. Vespignani, \emph{J. Econ. Interaction and Coordination} \textbf{2} 111 (2007).
\bibitem{2007_EPJB_Reichardt} J. Reichardt, D. R. White, \emph{Eur. Phys. J. B} \textbf{60} 217 (2007).
\bibitem{2008_AdvCS_Reyes} J. Reyes, S. Schiavo, G. Fagiolo, \emph{Adv. Complex Syst.} \textbf{11} 685 (2008).
\bibitem{2010_EconInq_Kali} R. Kali, J. Reyes, \emph{Econ. Inquiry} \textbf{48} 1072 (2010).
\bibitem{2010_NewJPhys_Garas} A. Garas, P. Argyrakis, C. Rozenblat, M. Tomassini, S. Havlin, \emph{New J. Phys.} \textbf{12} 113043 (2010).
\bibitem{2007_EPJB_Garlaschelli} D. Garlaschelli, T. Di Matteo, T. Aste, G. Caldarelli, M. I. Loffredo, \emph{Eur. Phys. J. B} \textbf{57} 159 (2007).
\bibitem{2009_ChinPL_Wen} D. Wen-Qi, S. Bo-Liang, \emph{Chin. Phys. Lett.} \textbf{26} 098902 (2009).
\bibitem{2002_JRes_Gleditsch} K. Gleditsch, \emph{J. Conflict Resolution} \textbf{46}712 (2002).
\bibitem{2004_PRE_Park} J. Park, M. E. J. Newman, \emph{Phys. Rev. E} \textbf{70} 066117 (2004).
\bibitem{2009_PRL_Garlaschelli} D. Garlaschelli, M. I. Loffredo, \emph{Phys. Rev. Lett.} \textbf{102} 038701 (2009).
\bibitem{1999_Science_Barabasi} A.-L. Barab\'{a}si, R. Albert, \emph{Science} \textbf{286} 509 (1999).
\bibitem{1957_PR_Jaynes} E. T. Jaynes, \emph{Phys. Rev.} \textbf{106} 620 (1957).
\bibitem{2002_PRL_Caldarelli} G. Caldarelli, A. Capocci, P. De Los Rios, M. A. Mu\~{n}oz, \emph{Phys. Rev. Lett.} \textbf{89} 258702 (2002).
\bibitem{2003_PRE_Boguna} M. Bogu\~{n}\'{a}, R. Pastor-Satorras, \emph{Phys. Rev. E} \textbf{68} 036112 (2003).
\bibitem{dataset} http://privatewww.essex.ac.uk/$\sim$ksg/exptradegdp.html.
\bibitem{PWT} A. Heston, R. Summers, B. Aten, World Table Version 6.1, Center for International Comparisons of Production, Income and Prices at the University of Pennsylvania (2002).
\bibitem{2004_PNAS_Barrat} A. Barrat, M. Barth\'{e}lemy, R. Pastor-Satorras, A. Vespignani, \emph{Proc. Natl Acad. Sci. USA} \textbf{101} 3747 (2004).
\bibitem{2004b_PRL_Garlaschelli} D. Garlaschelli, M. Loffredo, \emph{Phys. Rev. Lett.} \textbf{93} 268701 (2004).
\bibitem{2004_book_Feenstra} R. Feenstra, \emph{Advanced International Trade: Theory and Evidence } (Princeton Univ. Press, 2004). 
\bibitem{1979_AmEconRev_Anderson} J. E. Anderson, \emph{Amer. Econ. Rev.} \textbf{69} 106 (1979).
\bibitem{1985_RevEconStat_Bergstrand} J. H. Bergstrand, \emph{Rev. Econ. Statist. } \textbf{67} 474 (1985).
\bibitem{2010_JEcon_Fagiolo} G. Fagiolo, \emph{J. Econ. Interaction and Coordination} \textbf{5} 1 (2010).
\bibitem{2004_PT_Ruelle} D. Ruelle, \emph{Phys. Today} \textbf{57} 48 (2004).
\end{thebibliography}
\end{document}